\documentclass[aps,prb,10pt,twocolumn]{revtex4-1}
\usepackage{graphicx}
\usepackage{amssymb,amsmath,mathtools,textcomp}
\usepackage{bbm,enumerate}
\usepackage{hyperref}
\usepackage{txfonts}

\hypersetup{colorlinks, linkcolor={blue}, citecolor={blue}, urlcolor={blue}}

\newcommand{\rucl}{$\alpha$-RuCl$_3$}

\newcommand{\w}{\omega}
\newcommand{\wmax}{\omega_{\rm max}}
\newcommand{\wmin}{\omega_{\rm min}}

\newcommand{\iu}{\mathrm{i}}

\newcommand{\jeff}{j_{\rm eff}}
\newcommand{\hk}{Heisenberg-Kitaev}

\newcommand{\zzthree}{3f-zz}
\newcommand{\zzsix}{6f-zz}

\newcommand{\llangle}{\langle\!\langle}
\newcommand{\rrangle}{\rangle\!\rangle}

\newcommand{\lllangle}{\langle\!\langle\!\langle}
\newcommand{\rrrangle}{\rangle\!\rangle\!\rangle}

\graphicspath{{./}{./figs/}}

\date{May 25, 2020}

\begin{document}

\title{
Magnon dispersion
and dynamic spin response
in three-dimensional spin models for $\alpha$-RuCl$_3$
}

\author{Lukas Janssen}
\author{Stefan Koch}
\author{Matthias Vojta}
\affiliation{Institut f\"ur Theoretische Physik and W\"urzburg-Dresden Cluster of Excellence ct.qmat, Technische Universit\"at Dresden, 01062 Dresden, Germany}

\begin{abstract}
In the search for experimental realizations of bond-anisotropic Kitaev interactions and resulting spin-liquid phases, the layered magnet $\alpha$-RuCl$_3$ is a prime candidate. Its modelling typically involves Heisenberg, Kitaev, and symmetric off-diagonal $\Gamma$ interactions on the two-dimensional honeycomb lattice. However, recent neutron-scattering experiments point towards a sizeable magnetic interlayer coupling.
Here we study three-dimensional exchange models for $\alpha$-RuCl$_3$, for both possible $R\bar{3}$ and $C2/m$ crystal structures. We discuss the symmetry constraints on the interlayer couplings, construct minimal models, and use them to compute the magnetic mode dispersion and the dynamical spin structure factor, in both the zero-field zigzag phase and the paramagnetic high-field phase. Our predictions for the interlayer mode dispersion shall guide future experiments; they also call for a reevaluation of the quantitative model parameters relevant for $\alpha$-RuCl$_3$.
\end{abstract}

\maketitle

%%%%%%%%%%%%%%%%%%%%%%%%%%%%%%%%%%%%%%%%%%%%%%%%%%%%%%%%%%%%%%%%%%%%
%%%%%%%%%%%%%%%%%%%%%%%%%%%%%%%%%%%%%%%%%%%%%%%%%%%%%%%%%%%%%%%%%%%%
%%%%%%%%%%%%%%%%%%%%%%%%%%%%%%%%%%%%%%%%%%%%%%%%%%%%%%%%%%%%%%%%%%%%

\section{Introduction}

Mott-insulating magnets with strong spin-orbit coupling have become a major research field in condensed-matter physics.\cite{trebst2017,winter2017b,janssen2019,takagi2019} This has been partially triggered by Kitaev's construction \cite{kitaev2006} of a quantum spin liquid driven by bond-anisotropic exchange interactions on the honeycomb lattice, and by the subsequent proposal \cite{jackeli2009,chaloupka2010} to realize Kitaev interactions in layered honeycomb magnets with $\jeff=1/2$ moments.

Among the candidate materials, \rucl\ has received enormous interest. It displays low-temperature antiferromagnetic order of zigzag type, and this order can be suppressed by a moderate in-plane magnetic field.\cite{sears2015,johnson2015,leahy2017,baek2017,sears2017,wolter2017,zheng2017,hentrich2018} By now, the existence of a quantum spin-liquid phase in \rucl\ in a narrow window of magnetic fields is suggested by a number of experimental results, such as an excitation continuum in neutron scattering,\cite{banerjee2018,balz2019} a transition signature in magnetocaloric-effect measurements\cite{balz2019} and, most prominently, an approximately half-quantized thermal Hall conductivity,\cite{kasahara2018b,yokoi20} signifying the presence of a Majorana edge mode.

\rucl\ belongs to a family of layered van-der-Waals crystals, and due to the weak bonding between the layers its three-dimensional (3D) crystal structure appears to be fragile.
While it adopts\cite{johnson2015,cao2016} a monoclinic structure with space group $C2/m$ at room temperature, the low-temperature structure has been a matter of debate.\cite{winter2017b}
Here, three different structures have been reported, namely monoclinic $C2/m$, trigonal $P3_112$, and rhombohedral $R\bar 3$; they are distinguished by the pattern and sequence of the stacking of the honeycomb layers.\cite{fletcher1967,cao2016,park2016,kim2016} Experimentally, stacking faults appear frequently, which also significantly influence the magnetic properties: Early samples displayed two thermodynamic transitions at $T_{\mathrm N}=8$\,K and $14$\,K, while more recent higher-quality samples show a single transition at $7$\,K.\cite{sears2015,cao2016,banerjee2017}
For some recent samples, a structural phase transition was found\cite{kubota2015,glamazda2017,reschke2017,kelley2018a} around \mbox{100--150\,K}, with the refinement of the neutron-diffraction data consistent with the rhombohedral $R\bar 3$ structure at low temperature.\cite{park2016}

Most theoretical descriptions of the magnetism of \rucl\ have been restricted to planar exchange Hamiltonians, with the magnetic interlayer coupling assumed to be negligible. In contrast, recent inelastic neutron-scattering data \cite{balz2019} indicate a significant out-of-plane dispersion of magnetic excitations. This calls for a modelling of the relevant interlayer interactions and their consequences, which is lacking to our knowledge.

It is the purpose of this paper to close this gap. For the two most probable crystal structures $R\bar{3}$ and $C2/m$, we shall discuss symmetries and corresponding minimal models for the magnetic interlayer couplings. We then employ spin-wave theory to calculate the 3D magnetic mode dispersion and the dynamic spin structure factor, both in the zero-field zigzag phase as well as in the high-field phase. Comparing our results to experimental data, we obtain a consistent description of a variety of experimental data for an assumed $R\bar{3}$ crystal structure at low $T$. Our results provide concrete predictions for future experiments and will help constraining the model parameters relevant for \rucl: The sizeable interlayer coupling cannot be neglected when fitting experimental data, and consequently estimates for the intralayer couplings from previous modelling need to be revised.

The remainder of the paper is organized as follows:
In Sec.~\ref{sec:models}, we summarize the structural models put forward for \rucl\ and discuss the magnetic exchange Hamiltonians, with focus on the $R\bar{3}$ and $C2/m$ structures.
Section~\ref{sec:swt} discusses the application of spin-wave theory and illustrates the effect of interlayer by comparing the mode dispersion and the dynamic structure factor for a model with and without interlayer coupling.
In Sec.~\ref{sec:choice}, we use analytical results in the high-field phase together with available experimental data to derive constraints on the model parameters, which enable us to construct plausible parameters sets for 3D exchange models.
In Sec.~\ref{sec:res}, we then display numerical results for these constrained 3D models and discuss them vis-{\`a}-vis published experimental data.
A summary of our results, together with suggestions for future experiments, closes the paper.

%%%%%%%%%%%%%%%%%%%%%%%%%%%%%%%%%%%%%%%%%%%%%%%%%%%%%%%%%%%%%%%%%%%%
%%%%%%%%%%%%%%%%%%%%%%%%%%%%%%%%%%%%%%%%%%%%%%%%%%%%%%%%%%%%%%%%%%%%
%%%%%%%%%%%%%%%%%%%%%%%%%%%%%%%%%%%%%%%%%%%%%%%%%%%%%%%%%%%%%%%%%%%%

\begin{figure*}
\includegraphics[width=\textwidth]{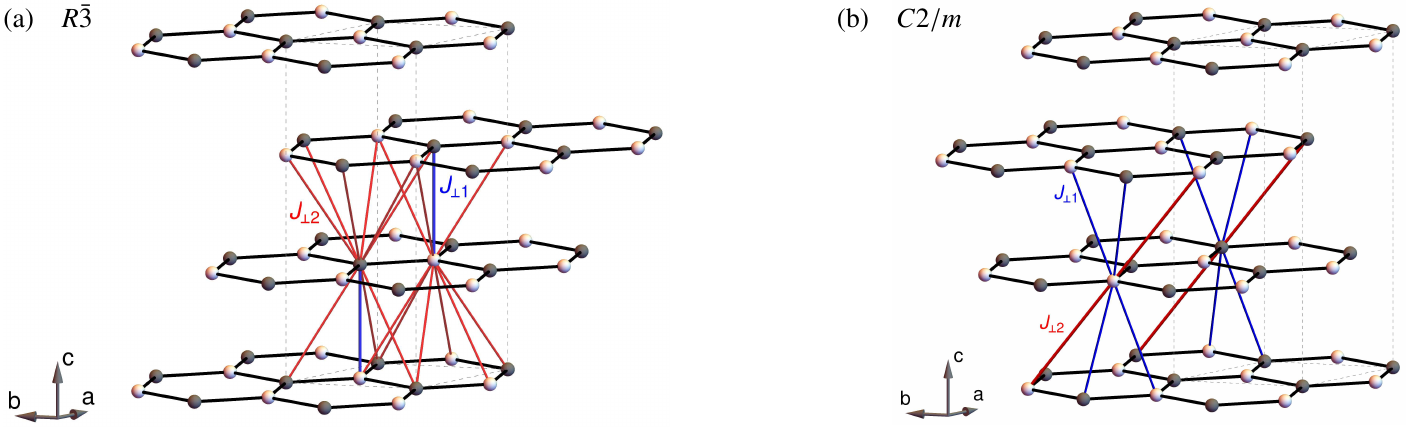}
\caption{
3D crystal structures of \rucl, showing the stacking of the honeycomb layers and the assumed interlayer interactions of the minimal models employed in this paper.
(a) $R\bar{3}$ structure,
(b) $C2/m$ structure.
In both cases, the conventional trigonal crystallographic unit cell (dashed) contains three RuCl$_3$ layers, and $\mathbf a$, $\mathbf b$, $\mathbf c$ show the directions of the basis vectors of this cell.
Each honeycomb layer consists of two sublattices, represented by black and white balls, respectively.
}
\label{fig:struct}
\end{figure*}

\section{Crystal structure and exchange models}
\label{sec:models}

\subsection{Intralayer exchange Hamiltonian}

The two-dimensional (2D) spin Hamiltonians proposed for {\rucl} are extensions\cite{rau2014} of the honeycomb-lattice {\hk} model originally introduced in Ref.~\onlinecite{chaloupka2010}.
%
%\cite{chaloupka2010}
%
For the purpose of this paper, we will consider the following in-plane exchange interactions
\begin{align} \label{eq:h0}
\mathcal{H}_0 & =
\sum_{n,\langle ij\rangle_\gamma} \Big[
J \vec S_{n,i} \cdot \vec S_{n,j}  + K S_{n,i}^\gamma S_{n,j}^\gamma
%
%\nonumber\\&~~~~~~~~~~~
%
+ \Gamma \left( S_{n,i}^\alpha S_{n,j}^\beta + S_{n,i}^\beta S_{n,j}^\alpha\right)
\Big]
\nonumber \\
&\quad + \sum_{n,\lllangle ij\rrrangle}
J_3 \vec S_{n,i} \cdot \vec S_{n,j}
- \vec h \cdot \sum_{n,i} \vec S_{n,i},
\end{align}
where $J$ and $J_3$ correspond to first- and third-neighbor Heisenberg couplings, while $K$ and $\Gamma$ are the first-neighbor Kitaev and symmetric off-diagonal couplings, respectively. $n$ is the layer index, and $\langle i j \rangle_\gamma$ denote first-neighbor $\gamma$ bonds, with $\gamma = x,y,z$. On $z$ bonds $(\alpha,\beta,\gamma) = (x,y,z)$, with cyclic permutation for $x$ and $y$ bonds. The uniform magnetic field is $\vec h \coloneqq g \mu_\mathrm{B} \mu_0 \vec H$, with $g$ the (possibly anisotropic) effective $g$ tensor and $\mu_\mathrm{B}$ the Bohr magneton.

Within each layer, we assume a perfect honeycomb structure, i.e., neglect possible trigonal distortions.\cite{agrestini2017} Then, $\mathcal{H}_0$ has a $C_3^\ast$ symmetry which combines a $120^\circ$ real-space rotation about a site with a spin rotation about the $[111]$ direction in spin space, exchanging $x\rightarrow y\rightarrow z\rightarrow x$. This symmetry also implies that the experimentally relevant zigzag state is threefold degenerate, with three symmetry-equivalent in-plane propagation directions.

For the extended \hk-$\Gamma$ model \eqref{eq:h0}, different parameter sets have been proposed to describe \rucl, based on either ab-initio modelling or on fits to experimental data, and we refer the reader to Ref.~\onlinecite{janssen2017} for an overview.
%
%\cite{janssen2017}
%
Guided by previous work,\cite{kim2016,winter2016,winter2017a,janssen2017,wangdong2017,winter2018,wolter2017,kelley2018b} we employ parameters where $K<0$ and $\Gamma>0$ are the dominant couplings, while both $J<0$ and $J_3>0$ are small, mainly acting to stabilize the zigzag phase. As will become clear below, the quantitative choice of the in-plane model parameters needs to be revisited upon including significant inter-layer interactions.

\subsection{$R\bar{3}$ structure and interlayer interactions}

\begin{figure}
\includegraphics[width=\linewidth]{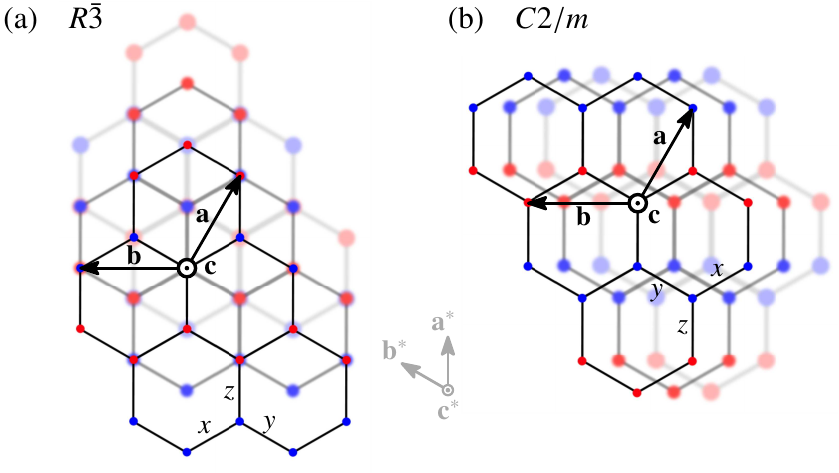}
\caption{
Top view of the 3D zigzag magnetic order with three-layer periodicity along the $c$ axis, showing three layers.
(a) $R\bar{3}$ structure,
(b) $C2/m$ structure.
Red and blue sites correspond to spin directions up and down, respectively.
$\mathbf a^*$, $\mathbf b^*$, $\mathbf c^*$ are the reciprocal lattice vectors in the conventional trigonal basis.
}
\label{fig:zz3}
\end{figure}

The rhombohedral structure with $R\bar{3}$ space group has a conventional crystallographic unit cell consisting of three honeycomb layers. The layers are stacked with a $(2\mathbf a + \mathbf b - \mathbf c)/3$ translation, Fig.~\ref{fig:struct}, such that the $C_3^\ast$ rotation symmetry is preserved and the honeycomb lattice is undistorted.
A second crystallographic domain, dubbed reverse-obverse twin,\cite{mcguire2015} can be obtained by a reflection in the $ab$ plane.
From neutron diffraction,\cite{cao2016,banerjee2016} it is known that the low-field zigzag phase in the samples with a single transition at $T_\mathrm{N} = 7\,$K exhibits a magnetic unit cell of three layers. This implies a stacked magnetic order as shown in Fig.~\ref{fig:zz3}(a), which we dub \zzthree.

We proceed by discussing a minimal model for magnetic interlayer couplings.
Each spin has one interlayer neighbor, which is located either right above or below it, depending on the sublattice index. This vertical spin-spin exchange interaction is compatible with the $C_3^*$ symmetry only for a Heisenberg coupling, denoted as $J_{\perp1}$ in Fig.~\ref{fig:struct}(a). Furthermore, each spin has nine next-nearest interlayer neighbors which fall into two classes (with six and three members, respectively) that are distinguished by the presence or absence of a nearest-neighbor intralayer bond in one of the participating layers. In the spirit of a minimal model, we will not distinguish between these different next-nearest interlayer neighbors, and assume Heisenberg interactions, $J_{\perp2}$, although spin-anisotropic interactions are symmetry-allowed here.
The interlayer part of the Hamiltonian thus reads
\begin{equation}
\label{eq:h1r3b}
\mathcal{H}_1^{R\bar{3}} = J_{\perp1} \! \sum_{\langle ni,mi\rangle}^1 \! \vec S_{n,i} \cdot \vec S_{m,i}
	+ J_{\perp2} \!\! \sum_{\llangle ni,mj\rrangle}^9 \!\! \vec S_{n,i} \cdot \vec S_{m,j}
\end{equation}
where the number above the summation symbol indicates the number of terms per spin.

Assuming that the nearest-neighbor interlayer bonds $J_{\perp1}$ dominate the interlayer exchange, the {\zzthree} configuration in Fig.~\ref{fig:zz3}(a) requires an \emph{antiferromagnetic} interlayer coupling, $J_{\perp1}>0$. We note that a ferromagnetic coupling would lead to a zigzag state with a unit cell of six layers (\zzsix); such a state is likely realized in \rucl\ in a small field window below the critical field,\cite{kelley2018b,balz2019} but we defer a detailed discussion of this to a future publication.\cite{balz2020}

The fact that all interlayer interactions are assumed to be of Heisenberg type implies that the direction of the zero-field ordered moment is unaffected by these interactions. As discussed extensively in Ref.~\onlinecite{janssen2017}, it is determined by the ratio of the anisotropic interactions $K$ and $\Gamma$.

\subsection{$C2/m$ structure and interlayer interactions}

The monoclinic structure with $C2/m$ space group also allows a conventional trigonal unit cell that consists of three honeycomb layers. Here, the layers are stacked with $(\mathbf b + \mathbf c)/3$ translation, such that, in contrast to $R\bar{3}$, the global $C_3^\ast$ rotation symmetry is broken, see Fig.~\ref{fig:struct}(b).
Other crystallographic domains can therefore be obtained by $C_3^*$ rotations.

In this structure, there are no vertical inter-layer bonds. According to the ab-initio analysis of Ref.~\onlinecite{kim2016},
%
%\cite{kim2016}
%
three types of inter-layer couplings are important and comparable in strength, namely first-neighbor bonds, such as those along $(\mathbf a+\mathbf c)/3$ and $(\mathbf b+\mathbf c)/3$, which we assume to be of equal strength, and second-neighbor bonds along $(2\mathbf b - \mathbf c)/3$. Although the form of these interactions is not symmetry-restricted, we confine ourselves to Heisenberg couplings $J_{\perp1}$ and $J_{\perp2}$, see Fig.~\ref{fig:struct}(b),
\begin{align}
\label{eq:h1c2m}
\mathcal{H}_1^{C2/m} &= J_{\perp1} \! \sum_{\langle ni,mi\rangle}^4 \! \vec S_{n,i} \cdot \vec S_{m,i}
	+ J_{\perp2} \!\! \sum_{\llangle ni,mj\rrangle}^2 \!\! \vec S_{n,i} \cdot \vec S_{m,j}.
\end{align}
Assuming again that the interlayer bonds obey $|J_{\perp1}| \gg |J_{\perp2}|$, realizing the {\zzthree} order here requires \emph{ferromagnetic} interlayer couplings, Fig.~\ref{fig:zz3}(b), as opposed to the antiferromagnetic couplings necessary in the $R\bar{3}$ structure.

Two remarks are in order:
(i) Since the $C2/m$ structure breaks the $C_3^\ast$ symmetry, the three propagation directions of the zigzag order are in general no longer degenerate. They remain, however, degenerate at the classical level within our model if we set $J_{\perp2}=0$.
(ii) If instead the interlayer coupling $J_{\perp2}$ dominates over $J_{\perp1}$, then an antiferromagnetic $J_{\perp2}$ may induce, depending on the sign of $J_{\perp1}$, either a {\zzsix} state propagating perpendicular to the $z$-bond or a {\zzthree} state propagating perpendicular to the $x$ or $y$ bond. We will not explore this option in detail.

%%%%%%%%%%%%%%%%%%%%%%%%%%%%%%%%%%%%%%%%%%%%%%%%%%%%%%%%%%%%%%%%%%%%
%%%%%%%%%%%%%%%%%%%%%%%%%%%%%%%%%%%%%%%%%%%%%%%%%%%%%%%%%%%%%%%%%%%%
%%%%%%%%%%%%%%%%%%%%%%%%%%%%%%%%%%%%%%%%%%%%%%%%%%%%%%%%%%%%%%%%%%%%

\section{Spin-wave theory and influence of interlayer coupling}
\label{sec:swt}

For the models of Sec.~\ref{sec:models}, we employ standard linear spin-wave theory for spins of size $S$ on the two different lattices.\cite{janssen2016, wolter2017,janssen2019}
We calculate the dynamic spin structure factor at $T=0$ according to
\begin{align}
\mathcal S(\mathbf q,\w) &= \sum_{\alpha} \int \mathrm d\tau\,\mathrm e^{\mathrm i\w\tau} \langle S^\alpha(\mathbf q, \tau) S^\alpha(-\mathbf q, 0)\rangle.
\end{align}
When specifying momenta, we will follow the conventions of Refs.~\onlinecite{banerjee2018,balz2019}
%
%\cite{banerjee2018,balz2019}
%
and use reciprocal-space coordinates $(H,K,L)$ in reciprocal lattice units, corresponding to an embedding trigonal unit cell. In this convention, the in-plane $\mathbf M$ point is located at $(H,K)=(0,1/2)$ while the in-plane $\mathbf K$ point is at $(-1/3,2/3)$. For the vertical direction, this convention implies that mode energies will be $L$-periodic with a period of $3$ in the {\zzthree} magnetic structure.

For discussing the parameter dependence of the results and relating them to experimental data, it is useful to define an overall energy scale $A$ and parameterize the couplings as $(J,K,\Gamma,J_3,J_{\perp1},J_{\perp2}) = A(\hat{J},\hat{K},\hat{\Gamma},\hat{J}_3,\hat{J}_{\perp1},\hat{J}_{\perp2})$. Similarly, we define the strength of the magnetic field as $|\vec{h}|=AS\hat{h}$.

\begin{figure*}
\includegraphics[width=\linewidth]{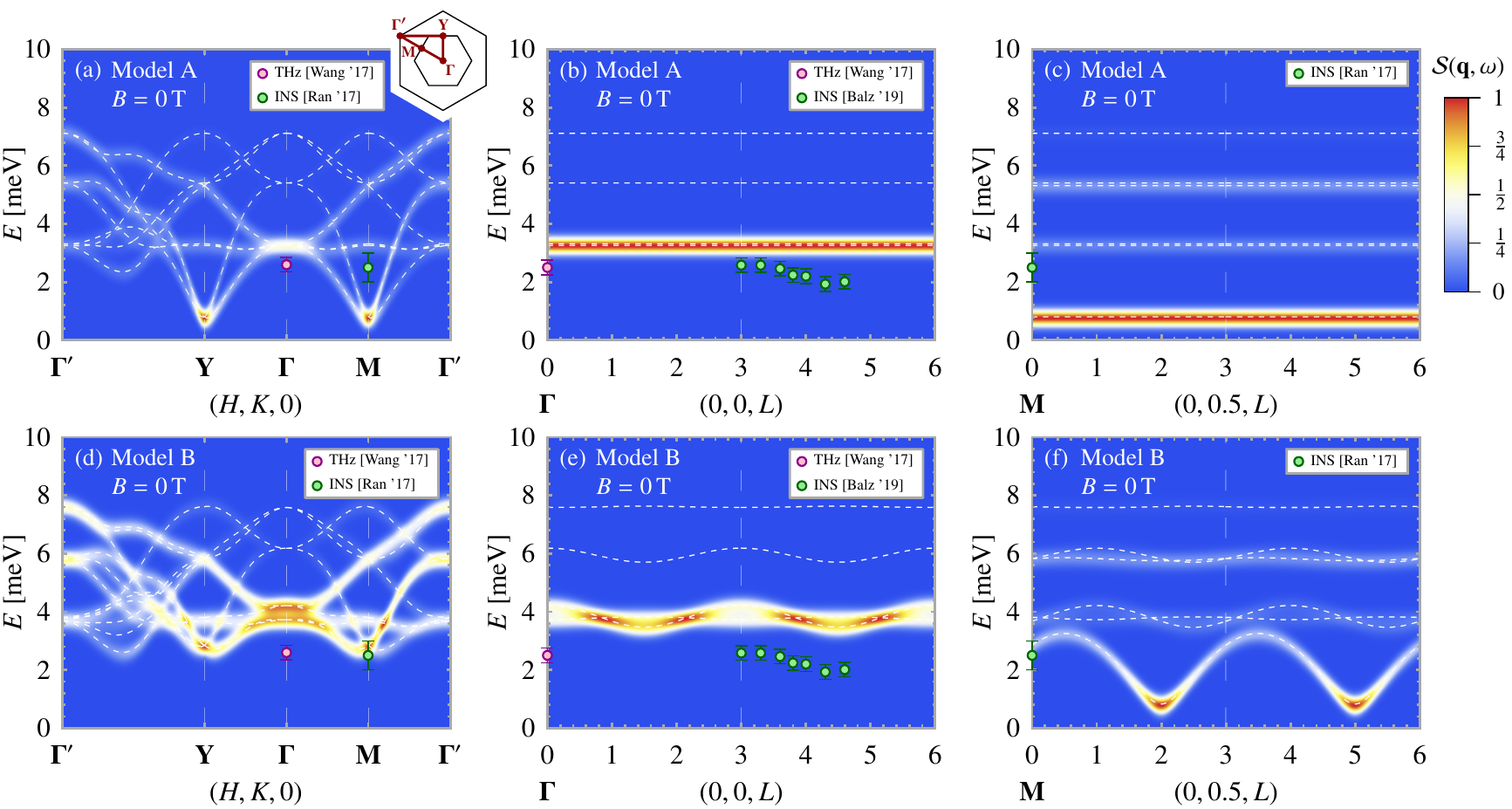}
\caption{
Dynamic spin structure factor $\mathcal{S}(\mathbf q,\w)$ (color-coded) and mode dispersion (dashed lines) at zero external field, calculated for
\mbox{(a-c)} the 2D Model A \cite{winter2017a} and
\mbox{(d-f)} the same model augmented by an interlayer coupling $J_{\perp1}$ of $1$\,meV in an assumed $R\bar{3}$ structure (Model B).
Different columns correspond to different paths in 3D momentum space:
(a,d) in-plane path as shown in panel (a) at $L=0$;
(b,e) vertical out-of-plane path at in-plane momentum $(0,0)$;
(c,f) vertical out-of-plane path at in-plane momentum $(0,0.5)$.
Symbols show mode energies extracted from THz spectroscopy (Ref.~\onlinecite{wang2017}) and
neutron scattering (INS, Refs.~\onlinecite{balz2019, ran2017}) measurements.
All panels involve an averaging over the three symmetry-equivalent zigzag domains.
}
\label{fig:dispAB_rb3}
\end{figure*}

\subsection{Phases}

In the high-field phase, the spin-wave expansion is performed about the polarized state. We will show results for magnetic fields along the two crystallographic in-plane directions perpendicular and parallel to a Ru-Ru bond, which correspond to the $(1, -2, 0)$ and $(1,0,0)$ directions in the reciprocal-space basis, respectively.\cite{note1}
In both cases, the magnetization in the high-field phase points along the field direction even in the presence of a finite $\Gamma$ term.\cite{janssen2017} For the $R\bar3$ ($C2/m$) structure, we work with a minimal two-site (four-site) unit cell, and the linear-spin-wave calculation amounts to performing a $4\times4$ ($8\times 8$) Bogoliubov transformation.
Calculational details are given in the Appendix.
A general introduction to spin-wave calculations in the context of Heisenberg-Kitaev-$\Gamma$ models can be found in the Appendix of Ref.~\onlinecite{janssen2019}.

In the zigzag phase, we work exclusively at zero field. The spin-wave expansion is performed about a {\zzthree} state, where the direction of the ordered moment is fixed by the ratio of $\Gamma$ and $K$, as explained in Sec.~VI of Ref.~\onlinecite{janssen2017}. In particular, for $\Gamma/|K|=1/2$ as used below, the magnetic moments $g\vec S$ point at an angle of $25^\circ$ out of plane if a $g$-factor anisotropy of $g_{ab}/g_c=1.77$ is used.\cite{kelley2018b}
As noted above, for the $R\bar{3}$ structure, there are three symmetry-equivalent propagation directions of the zigzag order. We perform the calculation of the spin structure factor for one of the three zigzag domains, obtain the result for the other domains by a $C_3^\ast$ rotation, and average the result over all three possible domains. For the $C2/m$ structure, this equivalence is violated, and we consider only the energetically favorable zigzag state with propagation direction perpendicular to the $z$ bond (assuming ferromagnetic interlayer couplings), without domain averaging.
Consequently, for both structures, the minimal magnetic unit cell contains four sites per layer and is periodically repeated in each layer, such that an $8\times8$ Bogoliubov transformation needs to be performed.

\subsection{Influence of interlayer coupling}

To illustrate the effect of the interlayer coupling on the excitation spectra, we start be presenting a comparison of the dynamic spin structure factor with and without interlayer coupling, keeping all other parameters fixed.
Fig.~\ref{fig:dispAB_rb3}(a-c) shows the spin-wave results for a strictly 2D parameter set taken from Ref.~\onlinecite{winter2017a}, $\hat{J}:\hat{K}:\hat{\Gamma}:\hat{J_3} = -0.1:-1:0.5:0.1$ with $A=5$\,meV. This model, which we dub Model A, has been frequently used in the recent literature. Panel (a) shows the dynamic structure factor for in-plane momenta; these data agree with Ref.~\onlinecite{winter2017a} where the corresponding neutron scattering intensity has been shown. Panels (b,c) correspond to momentum paths perpendicular to the plane.

In contrast, Fig.~\ref{fig:dispAB_rb3}(d-f) display the same information for Model B, which we obtain from Model A by adding an interlayer coupling $J_{\perp1}$ of $1$\,meV. This value has been chosen to approximately match the observed interlayer dispersion bandwidth.\cite{balz2019}
One sees that the inclusion of $J_{\perp1}$ increases the gap at $(H,K,L) = (0,0.5,0)$ significantly, while the gap at $(0,0,0)$ increases slightly and those at $(0,0,1.5)$ and $(0,0.5,2)$ are only changed minimally.

Fig.~\ref{fig:dispAB_rb3} illustrates that the agreement with experimentally measured mode energies is only moderate; in particular, the interlayer dispersion in panel (e) does not agree well with the one measured in neutron scattering.\cite{balz2019}
More seriously, while Model A with $g_{ab}=2.3$ (Ref.~\onlinecite{winter2018}) yields a classical critical field of around $11\,$T for $\vec H \parallel (1,0,0)$, in rough agreement with experiment,\cite{janssen2019,note2} Model B with the same $g_{ab}$ leads to $\mu_0 H_\mathrm{c} \simeq 19\,$T, which is far too large. The reason is that the interlayer coupling substantially stabilizes the zigzag phase, as will be further detailed below.

%%%%%%%%%%%%%%%%%%%%%%%%%%%%%%%%%%%%%%%%%%%%%%%%%%%%%%%%%%%%%%%%%%%%
%%%%%%%%%%%%%%%%%%%%%%%%%%%%%%%%%%%%%%%%%%%%%%%%%%%%%%%%%%%%%%%%%%%%
%%%%%%%%%%%%%%%%%%%%%%%%%%%%%%%%%%%%%%%%%%%%%%%%%%%%%%%%%%%%%%%%%%%%

\section{Constrained parameter choice}
\label{sec:choice}

Having seen that simply adding an interlayer coupling to previously used planar parameter sets for the model \eqref{eq:h0} leads to a sizeable mismatch between experiment and theory, in particular concerning the critical field, we now turn to a strategy which takes into account a larger set of experimental data in order to constrain the multi-dimensional model parameter set.
To this end, we find it useful to derive a few analytical results. In fact, for the high-field phase, one can determine the maxima and minima of the interlayer dispersion of the lowest magnon mode in closed forms, see Appendix.
We will use these together with the experimental information from Ref.~\onlinecite{balz2019} to guide the choice of model parameters.

\subsection{$R\bar{3}$ structure}

For the $R\bar{3}$ structure described by the model in Eq.~\eqref{eq:h1r3b}, we assume $J_{\perp1,2}>0$, which yields {\zzthree} order. For $\mathbf q =(0,0,L)$ and $\vec h \parallel (1,-2,0)$, the energy of the lowest mode in the high-field phase can be calculated as function of $L$.\cite{note3} It takes its extremal values at $L=0$ and $L=1.5$ in reciprocal lattice units. For $J_{\perp1,2}>0$, the maximum is at $L=0$ and reads
\begin{equation} \label{eq:omega-max-R3bar}
\wmax^2/(AS)^2 = \hat h(\hat h + 3 \hat\Gamma)
\end{equation}
which, remarkably, does not depend on the interlayer coupling. In the high-field limit, $\wmax\to |\vec{h}|$ as expected.
Similarly, the minimum energy $\wmin$ is taken at $L=1.5$ and evaluates to
\begin{equation} \label{eq:omega-min-R3bar}
\wmin^2/(AS)^2 = (\hat h - 2\hat{J}_{\perp1}-18\hat{J}_{\perp2})(\hat h - 2\hat{J}_{\perp1}-18\hat{J}_{\perp2}+3\hat\Gamma) \,.
\end{equation}
We can also compute the critical field for the disappearance of the zigzag order; it is given by
\begin{equation}
\label{eq:hcr3b}
\hat{h}_\mathrm{c} = 2\hat{J}+\hat{K}-\frac{\hat{\Gamma}}{2}+6\hat{J}_3+2\hat{J}_{\perp1}+10\hat{J}_{\perp2}
+\sqrt{\hat{K}^2-\hat{K}\hat{\Gamma}+\frac{9}{4}\hat{\Gamma}}
\end{equation}
for $\vec h\parallel (1,0,0)$; this is the direction for which the additional ordered phase found in Ref.~\onlinecite{kelley2018b} is either absent or very narrow. Eq.~\eqref{eq:hcr3b} underlines the collective role played by $J_3$, $J_{\perp1}$, and $J_{\perp2}$ in stabilizing the zigzag order, as all of them contribute to increase the critical field.

\begin{figure*}
\includegraphics[width=\linewidth]{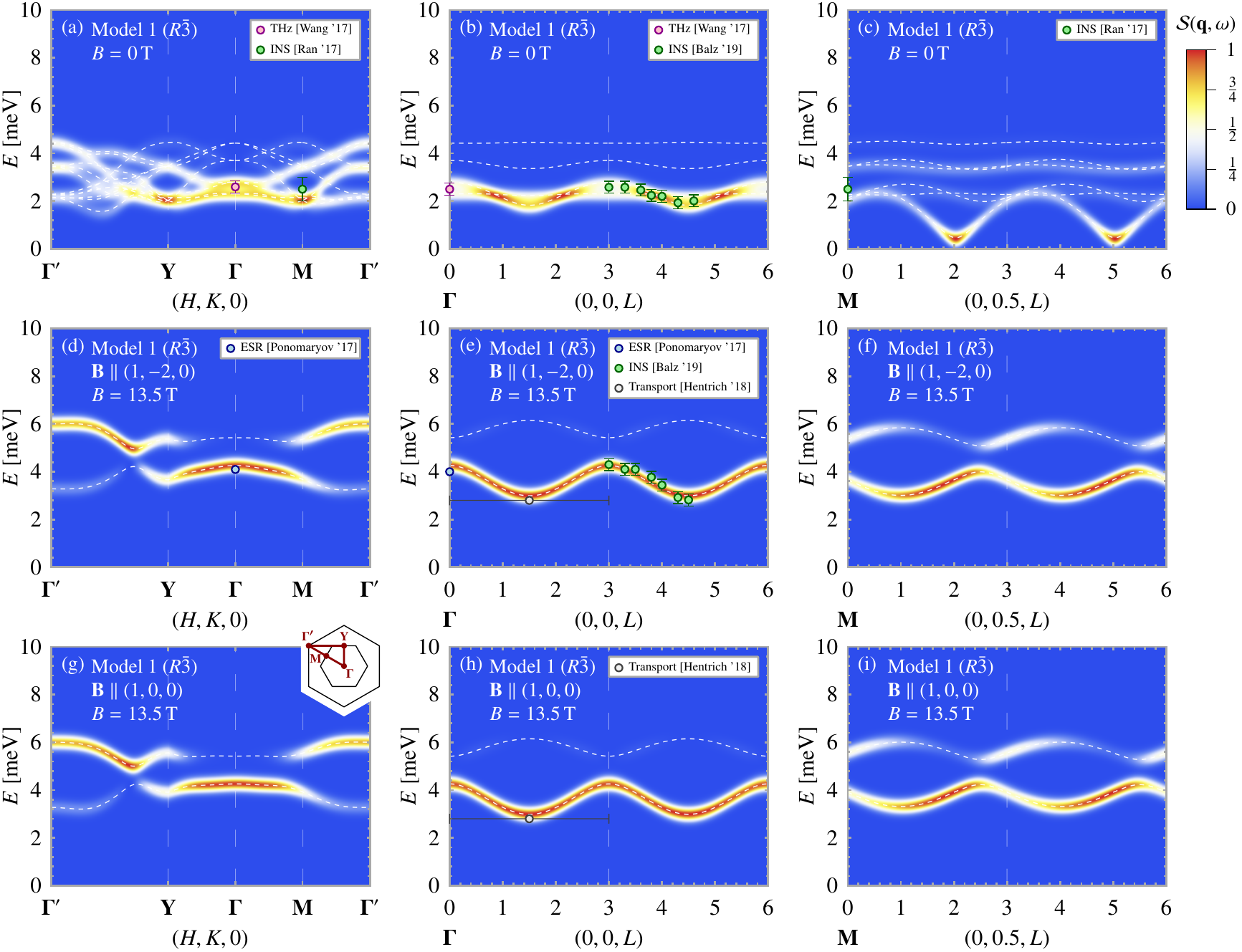}
\caption{
Dynamic spin structure factor $\mathcal{S}(\mathbf q,\w)$ (color-coded) and mode dispersion (dashed lines), calculated for Model 1 with $R\bar{3}$ crystal structure. Different rows correspond to different external parameters:
(a-c) Zero-field zigzag phase;
(d-f) high-field phase for field along $(1,-2,0)$ (perpendicular to a Ru-Ru bond);
(g-i) high-field phase for field along $(1,0,0)$ (parallel to a Ru-Ru bond).
Different columns correspond to different paths in 3D momentum space:
(a,d,g) in-plane path as shown in panel (g) at $L=0$;
(b,e,h) vertical out-of-plane path at in-plane momentum $(0,0)$;
(c,f,i) vertical out-of-plane path at in-plane momentum $(0,0.5)$.
Symbols show experimental mode energies extracted from THz spectroscopy (Ref.~\onlinecite{wang2017}), ESR (Ref.~\onlinecite{pono2017}),
neutron scattering (INS, Refs.~\onlinecite{balz2019, ran2017}), and thermal transport (Ref.~\onlinecite{hentrich2018}) measurements.
Panels (a-c) involve an averaging over the three symmetry-equivalent zigzag domains. The agreement with the experimental data is striking.
}
\label{fig:disp1_rb3}
\end{figure*}

\begin{figure*}
\includegraphics[width=\linewidth]{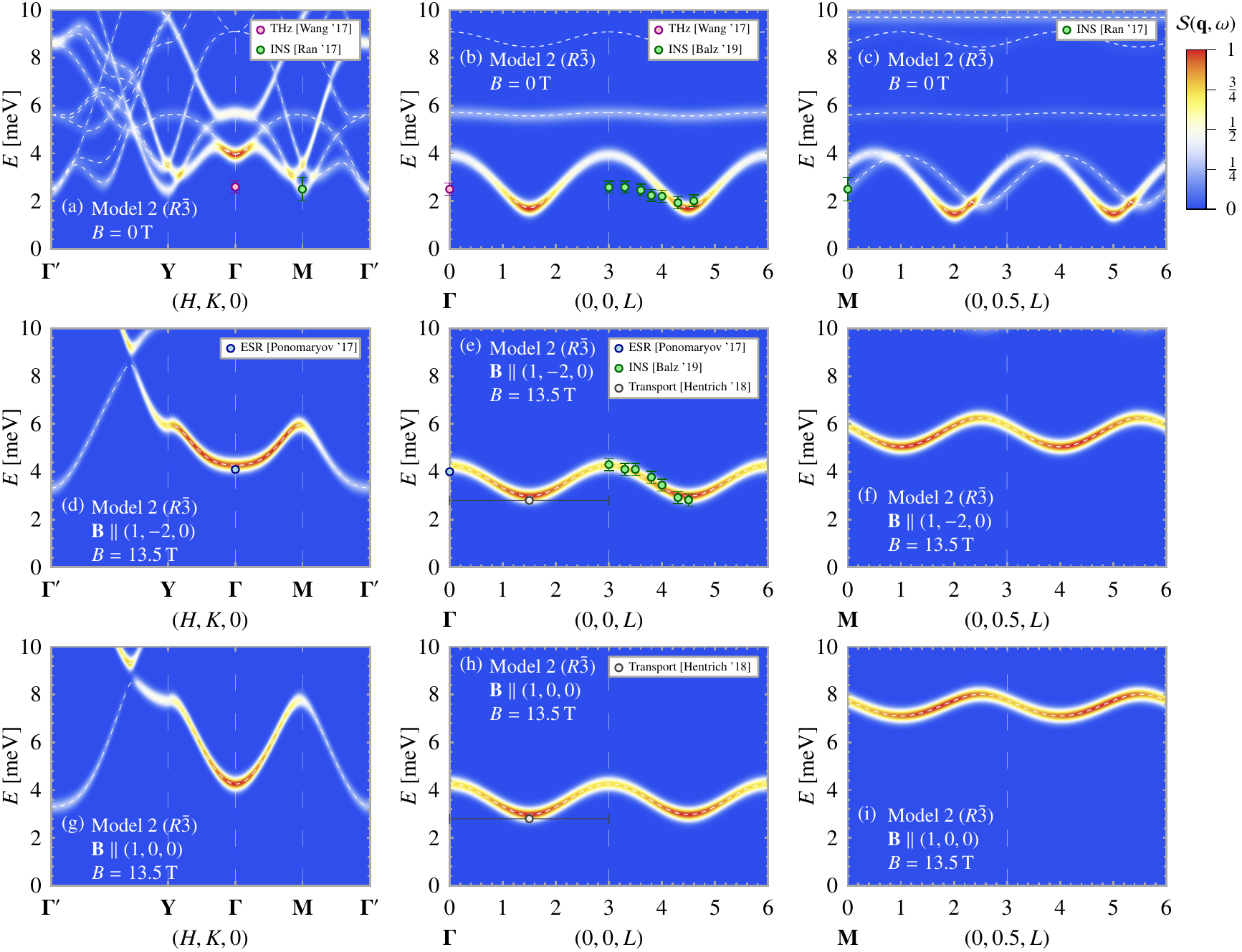}
\caption{
Same as Fig.~\ref{fig:disp1_rb3}, but now for Model 2 with $R\bar{3}$ crystal structure.
While this parameter set also reproduces the high-field mode energies (see text for details), its does not match quantitatively at zero field.
}
\label{fig:disp2_rb3}
\end{figure*}

\begin{table}[t]
\caption{Parameter sets for the spin models used in this paper: Model A is the 2D model of Refs.~\onlinecite{winter2017a, winter2018}; for Model B an interlayer coupling of $1$\,meV has been added. Models 1-3 arise from the considerations in Sec.~\ref{sec:choice}. The table also quotes the resulting critical field in units of $AS/(g\mu_\mathrm{B})$; for the constrained Models 1-3 $\mu_0 H_\mathrm{c} = \hat h_\mathrm{c} A S/(g\mu_\mathrm{B})$ evaluates to $7.6$\,T up to rounding errors.
}
\label{tab:par}
 \begin{tabular*}{\columnwidth}{@{\extracolsep{\fill} }lc | cccc | cc | cc | c}
   \hline\hline
   \# & Strct. & $\hat{J}$ & $\hat{K}$ & $\hat\Gamma$ & $\hat{J}_3$ & $\hat{J}_{\perp1}$ & $\hat{J}_{\perp2}$ & $A$\,[meV] & $g_{ab}$ & $\hat{h}_\mathrm{c}$
 \\ \hline
  A & p6m & $-0.1$ & $-1$ & $0.5$ & $0.1$ & n/a & n/a & $5$ & $2.3$ & $0.59$\\
  B & $R\bar{3}$ & $-0.1$ & $-1$ & $0.5$ & $0.1$ & $0.2$ & $0$ & $5$ & $2.3$ & $0.97$ \\
 \hline
  1 & $R\bar{3}$ & $-0.1$ & $-1$ & $0.5$ & $0.1$ & $0.3$ & $0.015$ & $2.8$ & $4.3$ & $1.34$ \\
  2 & $R\bar{3}$ & $-0.1$ & $-1$ & $0.5$ & $0.01$ & $0.08$ & $0.001$ & $10$ & $2.5$ & $0.22$ \\ \hline
  3 & $C2/m$     & $-0.1$ & $-1$ & $0.5$ & $0.1$ & $-0.05$ & $-0.05$ & $4$ & $2.7$ & $0.59$ \\
  \hline\hline
 \end{tabular*}
\end{table}

The experiment of Ref.~\onlinecite{balz2019} has determined the mode dispersion by inelastic neutron scattering, with the results $\wmax\approx4.3$\,meV and $\wmin\approx3$\,meV for $B=13.5$\,T applied along the $(1,-2,0)$ direction. Moreover, the experimental critical field is about $B_{\mathrm c}=7.6$\,T for the $(1,0,0)$ direction.\cite{kelley2018b} Together, this information can be used to constrain the model parameters; further constraints arise from fitting the zero-field interlayer dispersion, see below.
We note that accurate information on the in-plane spin-wave dispersion in {\rucl} is available neither at zero field nor at high fields,\cite{banerjee2016,banerjee2017,ran2017,banerjee2018,balz2019} leaving a considerable uncertainty in a conclusive determination of model parameters from mode dispersions.

\begin{figure*}
\includegraphics[width=\linewidth]{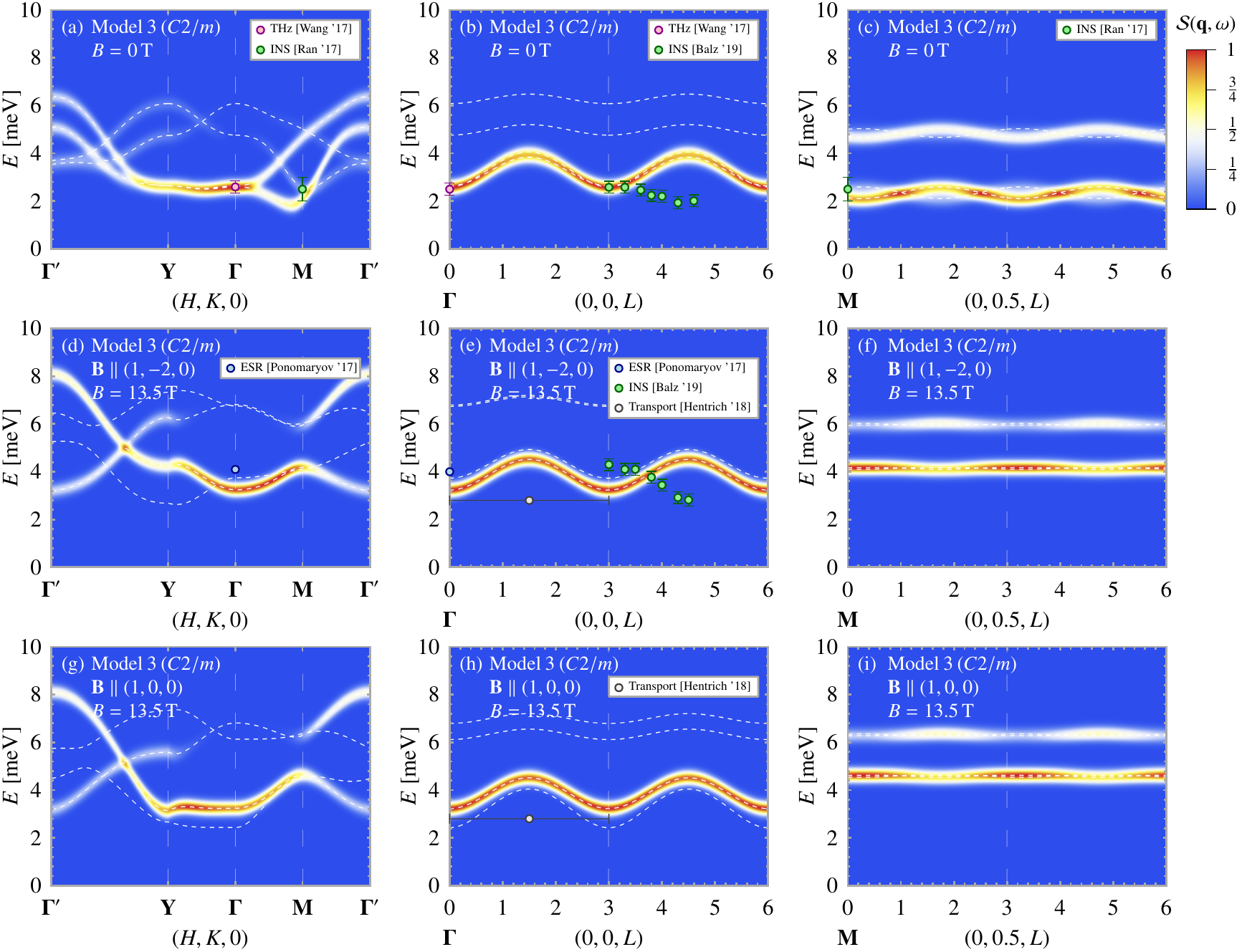}
\caption{
Same as Fig.~\ref{fig:disp1_rb3}, but now for Model 3, which has $C2/m$ crystal structure. Here, panels (a-c) have been calculated for the energetically favorable zigzag direction and do not involve domain averaging.
This model cannot reproduce the qualitative shape of the high-field dispersion; this generically applies to $C2/m$ models, which require a ferromagnetic interlayer coupling in order to stabilize threefold-periodic zigzag order in zero field.
}
\label{fig:disp3_c2m}
\end{figure*}

To build plausible sets of model parameters, we first fix the ratio of the in-plane nearest-neighbor couplings to $\hat{J}:\hat{K}:\hat{\Gamma} = -0.1:-1:0.5$, as obtained in Ref.~\onlinecite{winter2017a}.
%
%\cite{winter2017a}
%
We then choose ratios of $\hat J_3/\hat K$ and $\hat{J}_{\perp2}/\hat{J}_{\perp1}$. With these fixed, $A$ and $\hat{J}_{\perp1}$ are uniquely determined by matching $\wmin$ and $\wmax$  at $\hat h/\hat h_\mathrm{c} = 1.78$ with the experimental data at $13.5\,$T.
Finally, the value of $g_{ab}$ is determined by demanding that the critical field matches the experimental value, $\mu_0 H_\mathrm{c} = 7.6\,$T. Choosing $\hat J_3/\hat K$ and $\hat{J}_{\perp2}/\hat{J}_{\perp1}$ is not unique: To guide this, we monitor the numerical result at zero field (where we have no analytic solution) and try to match the vertical mode dispersion with the corresponding results of Ref.~\onlinecite{balz2019}.
%
%\cite{balz2019}
%
We also observe that the quality of this match depends only weakly on $\hat{J}_{\perp2}/\hat{J}_{\perp1}$.

Out of the family of possible parameter sets, we present results for two, which we dub Model 1 and Model 2, with the numerical parameter values shown in Table~\ref{tab:par}. In both cases, the in-plane parameters deviate substantially from the ones used before:\cite{janssen2017} Either $J_3$ is significantly smaller (Model 2), or all parameters are significantly smaller at the expense of a larger $g$ factor (Model 1). This mainly reflects the fact that the interlayer coupling, the sizable magnitude of which is dictated by the experimentally detected out-of-plane magnon dispersion, tends to stabilize the zigzag order and hence increases the critical field.

\subsection{$C2/m$ structure}

For the $C2/m$ structure described by the model in Eq.~\eqref{eq:h1c2m}, we now assume $J_{\perp1,2}<0$ in order to obtain {\zzthree} order.
The extremal energies of the lowest mode dispersion for $\mathbf q =(0,0,L)$ and $\vec h\parallel (1,-2,0)$ in the high-field phase are again taken at $L=0$ and $L=1.5$. However, for $J_{\perp1,2}<0$, the point $L=0$ now represent the dispersion minimum, with
\begin{equation} \label{eq:omega-min-C2m}
\wmin^2/(AS)^2 = (\hat h + 2\hat{J}_{\perp2})(\hat h +2\hat{J}_{\perp2} + 3\hat\Gamma)\,.
\end{equation}
Conversely, the maximum energy is now taken at $L=1.5$ and evaluates to
\begin{equation} \label{eq:omega-max-C2m}
\wmax^2/(AS)^2 = (\hat h - 8\hat{J}_{\perp1}-2\hat{J}_{\perp2})(\hat h - 8\hat{J}_{\perp1}-2\hat{J}_{\perp2}+3\hat\Gamma) \,.
\end{equation}
The critical field for the disappearance of the zigzag order is here given by
\begin{equation} \label{eq:hcC2m}
\hat{h}_\mathrm{c} = 2\hat{J}+\hat{K}-\frac{\hat{\Gamma}}{2}+6\hat{J}_3
+\sqrt{\hat{K}^2-\hat{K}\hat{\Gamma}+\frac{9}{4}\hat{\Gamma}}
\end{equation}
for $\vec h\parallel (1,0,0)$. Interestingly, this is independent of the interlayer coupling as in both the canted zigzag and high-field phases all spins coupled by $J_{\perp1,2}$ are aligned in parallel fashion.

Following a matching procedure similar to the one described above, we arrive at a parameter set which we dub Model 3, see Table~\ref{tab:par}. Here, the in-plane parameters are not very different from previous modelling,\cite{janssen2017} because the interlayer coupling is ferromagnetic and does not influence the critical field.

%%%%%%%%%%%%%%%%%%%%%%%%%%%%%%%%%%%%%%%%%%%%%%%%%%%%%%%%%%%%%%%%%%%%
%%%%%%%%%%%%%%%%%%%%%%%%%%%%%%%%%%%%%%%%%%%%%%%%%%%%%%%%%%%%%%%%%%%%
%%%%%%%%%%%%%%%%%%%%%%%%%%%%%%%%%%%%%%%%%%%%%%%%%%%%%%%%%%%%%%%%%%%%

\section{Results for constrained 3D models}
\label{sec:res}

We now turn to a discussion of the numerical results for the dynamic spin structure factor, shown in Figs.~\ref{fig:disp1_rb3}, \ref{fig:disp2_rb3}, and \ref{fig:disp3_c2m} for Models 1, 2, and 3, respectively. For comparison, we have added experimentally determined magnetic mode energies extracted from THz spectroscopy,\cite{wang2017} electron spin resonance (ESR),\cite{pono2017} inelastic neutron scattering (INS),\cite{ran2017, balz2019} and thermal transport\cite{hentrich2018} measurements.

For both Models 1 and 2, we obtain agreement with the high-field dispersion along $(0,0,L)$ as measured by INS; given the momentum-space location of minimum and maximum this agreement is achieved by construction. Model 1 also displays excellent agreement for the zero-field dispersion along $(0,0,L)$ as well as with the THz and ESR data; Model 2 performs inferior in this respect. We note that Model 1 has a rather small overall energy scale $A$ and a $g$ factor that is significantly larger than most\cite{kubota2015, agrestini2017, yadav2016} (but not all\cite{wang2017}) of the previous estimates.
For Model 2, the $g$ factor agrees well with the majority of previous results.

In contrast, Model 3 fails to match the high-field dispersion along $(0,0,L)$. This is because the momentum-space location of minimum and maximum are switched w.r.t. Models 1 and 2, because the interlayer coupling is \emph{ferro}magnetic here, $J_{\perp1,2}<0$.  We recall that this is required in order to stabilize the experimentally observed {\zzthree} magnetic order.

Given the mismatch visible in Fig.~\ref{fig:disp3_c2m}, we conclude that a $C2/m$ low-temperature crystal structure appears unlikely to be realized in \rucl. Instead, the assumption of an $R\bar{3}$ crystal structure with sizeable antiferromagnetic interlayer couplings leads to results consistent with experiment.

%%%%%%%%%%%%%%%%%%%%%%%%%%%%%%%%%%%%%%%%%%%%%%%%%%%%%%%%%%%%%%%%%%%%
%%%%%%%%%%%%%%%%%%%%%%%%%%%%%%%%%%%%%%%%%%%%%%%%%%%%%%%%%%%%%%%%%%%%
%%%%%%%%%%%%%%%%%%%%%%%%%%%%%%%%%%%%%%%%%%%%%%%%%%%%%%%%%%%%%%%%%%%%

\section{Summary}
\label{sec:summ}

Motivated by recent neutron-scattering results indicating a significant interlayer dispersion, we have discussed 3D spin models for the Kitaev material \rucl. We have considered two candidate crystal structures, $R\bar{3}$ and $C2/m$. For both, we have constructed minimal interlayer coupling models and determined the mode dispersion and dynamical spin structure factor using spin-wave theory, both in zero field and at high fields.

Our results show that the minimal models for the $C2/m$ structure cannot simultaneously reproduce the experimentally found zero-field magnetic structure and the form of the high-field interlayer dispersion. In contrast, the minimal models for $R\bar{3}$ can reproduce both, provided that the interlayer couplings are assumed to be sizeable and antiferromagnetic. In fact, in both Models 1 and 2 the interlayer coupling is of order $1$\,meV, as dictated by the experimentally observed\cite{balz2019} interlayer magnon bandwidth of $1.3$\,meV.

For simplicity, we have neglected spin anisotropies in the interlayer interactions. While those will be present (except for the vertical coupling in the $R\bar{3}$ structure) and will change the detailed quantitative fitting of experimental data, we expect our analysis to be semi-quantitatively robust concerning the magnitude of the interlayer couplings. Interlayer interaction anisotropies are likely to play a role for the intermediate ordered phase observed in {\rucl}; this will be explored in a forthcoming publication.\cite{balz2020}

Our analysis clearly shows that interlayer interactions cannot be neglected in {\rucl} when it comes to quantitative modelling, because these interactions substantially influence the stability of the zigzag phase.
We also note that sharp magnon modes will receive a broadening of the order of the interlayer magnon bandwidth, if scattering data are integrated over substantial ranges of the out-of-plane momentum.
More detailed neutron-scattering studies are therefore called for. In particular, the interlayer dispersion at high fields should be measured at various in-plane wavevectors, and ideally also various field directions, which will enable one to better discriminate between the parameter sets of Models 1 and 2.

%%%%%%%%%%%%%%%%%%%%%%%%%%%%%%%%%%%%%%%%%%%%%%%%%%%%%%%%%%%%%%%%%%%%

\acknowledgments

We thank C. Balz, B. B\"uchner, P. C\^{o}nsoli, W. Kr\"uger, P. Lampen-Kelley, R. Moessner, S. Nagler, S. Rachel, and A. Wolter
for illuminating discussions and collaboration on related work.
This work was funded by the Deutsche Forschungsgemeinschaft (DFG) through the Emmy Noether program (JA2306/4-1, project id 411750675), GRK 1621 (project id 129760637), SFB 1143 (project id 247310070), and the W\"urzburg-Dresden Cluster of Excellence on Complexity and Topology in Quantum Matter -- \textit{ct.qmat} (EXC 2147, project id 390858490).

%%%%%%%%%%%%%%%%%%%%%%%%%%%%%%%%%%%%%%%%%%%%%%%%%%%%%%%%%%%%%%%%%%%%

\appendix

\section{Spin-wave theory for 3D Heisenberg-Kitaev-$\Gamma$ models}

We employ a Holstein-Primakoff decomposition of the spin operator $\vec S_{n,j}$ at the $j$-th site in the $n$-th layer,
\begin{align} \label{eq:Holstein-Primakoff-decomposition}
	\vec S_{n,j} & = (S - a^\dagger_{n,j} a_{n,j}) \vec n_{n,j} 
	+ \sqrt{\frac{S}{2}} (a^\dagger_{n,j} + a_{n,j}) \vec e_{n,j}
	\nonumber \\ & \quad
	+ i \sqrt{\frac{S}{2}} (a_{n,j}^\dagger - a_{n,j}) (\vec n_{n,j} \times \vec e_{n,j}) 
	+ \mathcal{O}(1/\sqrt{S}),
\end{align}
where $\vec n_{n,j}$ denotes the spin direction in the classical limit ($S \to \infty$) and $\vec e_{n,j}$ represents an (arbitrary) unit vector in the plane perpendicular to $\vec n_{n,j}$.
In the high-field phase and for in-plane magnetic fields $\vec h \perp \mathbf c$, the classical spins point along the direction of the field, $\vec n \equiv \vec n_{n,j} \propto \vec h$, see Ref.~\onlinecite{janssen2017}.
In each phase, the Heisenberg-Kitaev-$\Gamma$ Hamiltonian can then be written as\cite{janssen2019}
\begin{align} \label{eq:Hamiltonian-SWT}
	\mathcal H_0 + \mathcal H_1 & = 
	S^2 \varepsilon_\text{cl} + \frac{S}{2} \sum_{\mathbf q}
	\begin{pmatrix}
		\vec \alpha_{\mathbf q} \\
		\vec \alpha_{-\mathbf q}^*
	\end{pmatrix}^\dagger
	\begin{pmatrix}
		K(\mathbf q) & \Delta^\dagger(\mathbf q) \\
		\Delta(\mathbf q) & K^\top(-\mathbf q) 
	\end{pmatrix}
	\begin{pmatrix}
		\vec \alpha_{\mathbf q} \\
		\vec \alpha_{-\mathbf q}^*
	\end{pmatrix}
	\nonumber \\ & \quad
	+ \mathcal O(\sqrt{S}),
\end{align}
where $S^2 \varepsilon_\text{cl}$ denotes the classical ground-state energy and $\vec \alpha_\mathbf{q} \equiv \left(\alpha_{\mathbf q s} \right)_{s = 1,\dots,M}$ with
\begin{align}
	\alpha_{\mathbf q s}
	= \sqrt{\frac{M}{N}} \sum_{(n,j) \in \text{$s$-th sublattice}} \mathrm{e}^{- \mathrm{i} \mathbf q \cdot \mathbf R_{n,j} } a_{n,j},
	\quad s=1,\dots,M,
\end{align}
and $\vec \alpha _{\mathbf q}^* \equiv \left(\vec \alpha_{\mathbf q}^\top\right)^\dagger$ are the vectors of magnon annihilation and creation operators, with $M$ the number of sites in the magnetic unit cell, $N$ the total number of sites, and $\mathbf R_{n,j}$ the position vector of the $j$-th site in the $n$-th layer.
The $M\times M$ matrices $K(\mathbf q)$ and $\Delta(\mathbf q)$ depend on the model and the particular phase. Exemplary results for the high-field phase of the $R\bar3$ and $C2/m$ models are given below.
The magnon spectrum $\{ \omega_{\mathbf q}^{(s)}\}_{s=1,\dots,M}$ is then obtained by a Bogoliubov transformation, which essentially amounts to solving the eigenvalue equation
\begin{align} \label{eq:eigenvalue}
	\begin{pmatrix}
		K(\mathbf q) & \Delta^\dagger(\mathbf q) \\
		-\Delta(\mathbf q) & -K^\top(-\mathbf q) 
	\end{pmatrix}
	\begin{pmatrix}
		\vec u_\mathbf{q}^{(s)} \\
		\vec v_{-\mathbf{q}}^{*(s)} \\		
	\end{pmatrix}
	=
	\omega_{\mathbf q}^{(s)}
	\begin{pmatrix}
		\vec u_\mathbf{q}^{(s)} \\
		\vec v_{-\mathbf{q}}^{*(s)} \\		
	\end{pmatrix},
\end{align}
as explained in detail in Ref.~\onlinecite{janssen2019}. Here, $s=1,\dots,M$ labels the different magnon bands in the Brillouin zone and $\left(\vec u_\mathbf{q}^{(s)}, \vec v_{-\mathbf{q}}^{*(s)}\right)^\top$ correspond to the respective eigenvectors.

Assuming that the eigenvectors are normalized according to 
$\vec{u}_\mathbf{q}^{(s)\dagger} \vec u_\mathbf{q}^{(s)} - \vec{v}_{-\mathbf{q}}^{*(s)\dagger} \vec{v}_{-\mathbf{q}}^{*(s)} = 1$, 
the dynamic spin structure factor is given by\cite{janssen2019}
\begin{align}
	\mathcal S(\mathbf q,\omega) & = \frac{S}{2} \sum_{s}^M \sum_{m,m'}^M 2\pi \delta(\omega - \omega_{\mathbf q}^{(s)})
	\left[ u_{\mathbf q m}^{(s)} u_{\mathbf q m'}^{*(n)} + v_{-\mathbf q m}^{*(s)} v_{-\mathbf q m'}^{(n)} \right] 
	\nonumber \\ & \quad
	+ \mathcal O(\delta(\omega), S^0).
\end{align}

\subsection{High-field phase on $R\bar3$ lattice}

In the $R\bar3$ structure, the high-field phase allows a minimal magnetic unit cell with $M=2$ sites and magnetic unit-cell vectors
\begin{align}
	\mathbf a_1^{R\bar3} & =
	\begin{pmatrix}
	3a_0/2 \\ -\sqrt{3/2}a_0 \\ 0
	\end{pmatrix},
	&
	\mathbf a_2^{R\bar3} & =
	\begin{pmatrix}
	3a_0/2 \\ \sqrt{3/2}a_0 \\ 0
	\end{pmatrix},
	&	
	\mathbf a_3^{R\bar3} & =
	\begin{pmatrix}
	a_0 \\ 0 \\ d_\perp
	\end{pmatrix},
\end{align}
using a Cartesian coordinate system with first (third) axis along the $(H,K,L) = (1,0,0)$ [$(H,K,L) = (0,0,1)$] direction, and 
where $a_0 \simeq 3.45(1)\,${\AA} is the distance between neighboring Ru ions of the same layer and $d_\perp \simeq 5.67(3)\,${\AA} the distance between neighboring layers.\cite{johnson2015,cao2016,park2016}
In terms of the intralayer nearest-neighbor vectors on the $x$, $y$, and $z$ bonds
\begin{align} \label{eq:NN-vectors}
	\boldsymbol\delta_\perp^x & =
	\begin{pmatrix}
		-a_0/2 \\ \sqrt{3}a_0/2 \\ 0
	\end{pmatrix}, & 
	\boldsymbol\delta_\perp^y & = 
	\begin{pmatrix}
		-a_0/2 \\ -\sqrt{3}a_0/2 \\ 0
	\end{pmatrix}, & 
	\boldsymbol\delta_\perp^z & =
	\begin{pmatrix}
		a_0 \\ 0 \\ 0
	\end{pmatrix},
\end{align}
and the interlayer nearest-neighbor vector
$\boldsymbol \delta_\perp = d_\perp(0,0,1)^\top$,
the $2\times2$ block occurring on the diagonal of the matrix in Eq.~\eqref{eq:eigenvalue} reads
\begin{multline}
	K(\mathbf q) = 
	\\
	\begin{pmatrix}
		\varepsilon_0 + \lambda_{0+}^\perp(\mathbf q) + \lambda_{0+}^\perp(-\mathbf q)&
		\lambda_0(\mathbf q) + \varepsilon_0^\perp(-\mathbf q) + \lambda_{0-}^\perp(\mathbf q) \\
		\lambda_0(-\mathbf q) + \varepsilon_0^\perp(\mathbf q) + \lambda_{0-}^\perp(-\mathbf q) &
		\varepsilon_0 + \lambda_{0+}^\perp(\mathbf q) + \lambda_{0+}^\perp(-\mathbf q)
	\end{pmatrix}
\end{multline}
with the on-site contributions as
\begin{align}
	\varepsilon_0 & = - 3 J - 3 J_3 - K - 2 \Gamma \sum_\gamma (\vec n \cdot \vec e_\alpha) (\vec n \cdot \vec e_\beta)
	- J_{\perp1} - 9 J_{\perp2}
	\nonumber \\ & \quad 
	+ h/S, 
\end{align}
the intralayer contributions as
\begin{align}
	\lambda_0(\mathbf q) & = \sum_\gamma \mathrm e^{\mathrm{i}\mathbf q\cdot \boldsymbol\delta_\gamma}  \Biggl\{ 
	J
	+ \frac{K}{2} \left[(\vec e \cdot \vec e_\gamma)^2 + \left((\vec n \times \vec e) \cdot \vec e_\gamma\right)^2 \right]
	\nonumber \\ & \quad
	+ \Gamma \left[(\vec e \cdot \vec e_\alpha)(\vec e \cdot \vec e_\beta) + ((\vec n \times \vec e) \cdot \vec e_\alpha) \left((\vec n \times \vec e) \cdot \vec e_\beta\right) \right]
	\Biggr\}
	\nonumber \\ & \quad 
	+ \sum_\gamma   \mathrm e^{-2\mathrm{i} \mathbf q\cdot\boldsymbol\delta_\gamma} J_3,
\end{align}
and the interlayer contributions as
\begin{align}
	\varepsilon_0^\perp(\mathbf q) & = J_{\perp1}\,\mathrm{e}^{\mathrm i \mathbf q \cdot \boldsymbol \delta_{\perp}},
	&
	\lambda_{0\pm}^\perp(\mathbf q) & = 
	J_{\perp2} \sum_\gamma \mathrm{e}^{\mathrm{i} \mathbf q \cdot (\pm \boldsymbol{\delta}_\gamma + \boldsymbol{\delta}_\perp)}.
\end{align}
In the above equations, $(\alpha,\beta,\gamma)$ is a permutation of $(x,y,z)$, the unit vectors $\vec n \equiv \vec n_{n,j}$ and $\vec e \equiv \vec e_{n,j}$ have been introduced in Eq.~\eqref{eq:Holstein-Primakoff-decomposition}, and $\vec e_x$, $\vec e_y$, and $\vec e_z$ correspond to the cubic spin-space vectors.
The off-diagonal $2\times2$ block of the matrix in Eq.~\eqref{eq:eigenvalue} is independent of the (isotropic) interlayer couplings and reads
\begin{align}
	\Delta(\mathbf q) = \begin{pmatrix}
	0 & \lambda_1(\mathbf q)\\
	\lambda_1(-\mathbf q) & 0
	\end{pmatrix}
\end{align}
with 
\begin{align}
		\lambda_1(\mathbf q) & = \sum_\gamma  \mathrm e^{\iu\mathbf q\cdot\boldsymbol\delta_\gamma} \Biggl\{
	\frac{K}{2} \left[(\vec e \cdot \vec e_\gamma) - \iu \left((\vec n \times \vec e) \cdot \vec e_\gamma\right) \right]^2
	\nonumber \\ & \quad
	+ \Gamma \Bigl[(\vec e \cdot \vec e_\alpha)(\vec e \cdot \vec e_\beta) - ((\vec n \times \vec e) \cdot \vec e_\alpha) \left((\vec n \times \vec e) \cdot \vec e_\beta\right) 
	\nonumber \\ & \quad
		- \iu (\vec e \cdot \vec e_\alpha)\left((\vec n \times \vec e) \cdot \vec e_\beta\right) - \iu (\vec e \cdot \vec e_\beta)\left((\vec n \times \vec e) \cdot \vec e_\alpha\right) 
	\Bigr]
	\Biggr\}.
\end{align}
For $J_{\perp1} = J_{\perp2} = 0$, the resulting Hamiltonian agrees with the previous reports for the two-dimensional Heisenberg-Kitaev-$\Gamma$ models.\cite{janssen2016,wolter2017,janssen2019}

For various high-symmetry wavevectors, including $\mathbf q = (0,0,0)$, $\mathbf q = (0,0,1.5)$, and $\mathbf q = (0.5,0,0)$ in the trigonal $(H,K,L)$ notation, the eigenvalue equation \eqref{eq:eigenvalue} can be solved in closed form. This leads to the analytical formulae for the maxima and the minima of the dispersion [Eqs.~\eqref{eq:omega-max-R3bar} and \eqref{eq:omega-min-R3bar}] for $\mathbf h \parallel (1,-2,0)$, and the critical field [Eq.~\eqref{eq:hcr3b}] for $\mathbf h \parallel (1,0,0)$, the latter being obtained by demanding that the magnon gap vanishes at the ordering wavevector for $\hat h \to \hat h_\mathrm{c}+$.

\subsection{High-field phase on $C2/m$ lattice}

In the $C2/m$ structure, the minimal magnetic unit cell in the high-field phase requires $M=4$ sites. We use the basis vectors
\begin{align}
	\mathbf a_1^{C2/m} & =
	\begin{pmatrix}
	3a_0 \\ 0 \\ 0
	\end{pmatrix},
	&
	\mathbf a_2^{C2/m} & =
	\begin{pmatrix}
	0 \\ \sqrt{3}a_0 \\ 0
	\end{pmatrix},
	&	
	\mathbf a_3^{C2/m} & =
	\begin{pmatrix}
	0 \\ \sqrt{3}a_0/3 \\ d_\perp
	\end{pmatrix},
\end{align}
within the same Cartesian coordinate system and $a_0$, $d_\perp$ as before.
The intralayer nearest-neighbor vectors $\boldsymbol\delta_x, \boldsymbol\delta_y, \boldsymbol\delta_z$ are the same as in Eq.~\eqref{eq:NN-vectors}; by contrast, the  interlayer nearest-neighbor vectors are now
\begin{align}
	\boldsymbol\delta_\perp^x & =
	\begin{pmatrix}
		-a_0/2 \\ -\sqrt{3}a_0/6 \\ d_\perp
	\end{pmatrix}, & 
	\boldsymbol\delta_\perp^y & = 
	\begin{pmatrix}
		a_0/2 \\ -\sqrt{3}a_0/6 \\ d_\perp
	\end{pmatrix}, & 
	\boldsymbol\delta_\perp^z & =
	\begin{pmatrix}
		0 \\ \sqrt{3}a_0/3 \\ d_\perp
	\end{pmatrix},
\end{align}
and the next-nearest-neighbor vector for the interlayer coupling $J_{\perp2}$ is $\boldsymbol \delta_{\perp2}^z = (0,-2\sqrt{3}a_0/3,d_\perp)^\top$.

\begin{widetext}

The $4\times4$ block occurring in the diagonal of the matrix in Eq.~\eqref{eq:eigenvalue} then reads
\begin{align}
	K(\mathbf q) = \begin{pmatrix}
		\varepsilon_0 + \lambda_{0}^{\perp z}(\mathbf q) + \lambda_{0}^{\perp z}(-\mathbf q) &
		\lambda_0^{x,y}(-\mathbf q) + \lambda_0^{\perp x}(-\mathbf q) + \lambda_{0}^{\perp y}(\mathbf q) &
		0 &
		\lambda_0^z(-\mathbf q) + \lambda_0^3(\mathbf q) \\
		\lambda_0^{x,y}(\mathbf q) + \lambda_0^{\perp x}(\mathbf q) + \lambda_{0}^{\perp y}(-\mathbf q) &
		\varepsilon_0 + \lambda_{0}^{\perp z}(\mathbf q) + \lambda_{0}^{\perp z}(-\mathbf q) &
		\lambda_0^z(\mathbf q) + \lambda_0^3(-\mathbf q) &
		0 \\
		0 &
		\lambda_0^z(-\mathbf q) + \lambda_0^3(\mathbf q) &
		\varepsilon_0 + \lambda_{0}^{\perp z}(\mathbf q) + \lambda_{0}^{\perp z}(-\mathbf q) &
		\lambda_0^{x,y}(-\mathbf q) + \lambda_0^{\perp x}(-\mathbf q) + \lambda_{0}^{\perp y}(\mathbf q) \\
		\lambda_0^z(\mathbf q) + \lambda_0^3(-\mathbf q) &
		0 &
		\lambda_0^{x,y}(\mathbf q) + \lambda_0^{\perp x}(\mathbf q) + \lambda_{0}^{\perp y}(-\mathbf q) &
		\varepsilon_0 + \lambda_{0}^{\perp z}(\mathbf q) + \lambda_{0}^{\perp z}(-\mathbf q)
	\end{pmatrix},
\end{align}
\end{widetext}
with the on-site contributions as
\begin{align}
	\varepsilon_0 & = - 3 J - 3 J_3 - K - 2 \Gamma \sum_{\gamma=x,y,z} (\vec n \cdot \vec e_\alpha) (\vec n \cdot \vec e_\beta)
	- 4J_{\perp1} - 2 J_{\perp2}
	\nonumber \\ & \quad 
	+ h/S, 
\end{align}
the intralayer contributions as
\begin{align}
	\lambda_0^{x,y}(\mathbf q) & = \sum_{\gamma=x,y} \mathrm e^{\mathrm{i}\mathbf q\cdot \boldsymbol\delta_\gamma}  \Biggl\{ 
	J
	+ \frac{K}{2} \left[(\vec e \cdot \vec e_\gamma)^2 + \left((\vec n \times \vec e) \cdot \vec e_\gamma\right)^2 \right]
	\nonumber \\ & \quad
	+ \Gamma \left[(\vec e \cdot \vec e_\alpha)(\vec e \cdot \vec e_\beta) + ((\vec n \times \vec e) \cdot \vec e_\alpha) \left((\vec n \times \vec e) \cdot \vec e_\beta\right) \right]
	\Biggr\}, \displaybreak[1] \\
		\lambda_0^{z}(\mathbf q) & = \mathrm e^{\mathrm{i}\mathbf q\cdot \boldsymbol\delta_z}  \Biggl\{ 
	J
	+ \frac{K}{2} \left[(\vec e \cdot \vec e_z)^2 + \left((\vec n \times \vec e) \cdot \vec e_z\right)^2 \right]
	\nonumber \\ & \quad
	+ \Gamma \left[(\vec e \cdot \vec e_x)(\vec e \cdot \vec e_y) + ((\vec n \times \vec e) \cdot \vec e_x) \left((\vec n \times \vec e) \cdot \vec e_y\right) \right]
	\Biggr\}, \displaybreak[1] \\
	\lambda_0^{3}(\mathbf q) & = J_3 \sum_{\gamma=x,y,z} \mathrm{e}^{2\mathrm{i} \mathbf q \cdot \boldsymbol{\delta}_\gamma},
\end{align}
and the interlayer contributions as
\begin{align}
%	\lambda_{0}^{\perp x}(\mathbf q) & = 
%%	
%	J_{\perp1} \mathrm{e}^{\mathrm{i} \mathbf q \cdot \boldsymbol{\delta}_\perp^x},
%%
%	&
%%
%	\lambda_{0}^{\perp y}(\mathbf q) & = 
%%	
%	J_{\perp1} \mathrm{e}^{\mathrm{i} \mathbf q \cdot \boldsymbol{\delta}_\perp^y},
%%
%	&
%%
%	\lambda_{0}^{\perp z}(\mathbf q) & = 
%%	
%	J_{\perp1} \mathrm{e}^{\mathrm{i} \mathbf q \cdot \boldsymbol{\delta}_\perp^z}
%	+ J_{\perp2} \mathrm{e}^{\mathrm{i} \mathbf q \cdot \boldsymbol{\delta}_{\perp2}^z}.
%
	\lambda_{0}^{\perp \gamma}(\mathbf q) & = 
	J_{\perp1} \mathrm{e}^{\mathrm{i} \mathbf q \cdot \boldsymbol{\delta}_\perp^\gamma}
	+ J_{\perp2} \mathrm{e}^{\mathrm{i} \mathbf q \cdot \boldsymbol{\delta}_{\perp2}^z} \delta_{\gamma,z}.
\end{align} 
As in the case of the $R\bar3$ structure, the (isotropic) interlayer interactions do not contribute to the off-diagonal block of the matrix in Eq.~\eqref{eq:eigenvalue}, which reads
\begin{align}
	\Delta(\mathbf q) = \begin{pmatrix}
	0 & \lambda_1^{x,y}(-\mathbf q) & 0& \lambda_1^{z}(-\mathbf q) \\
	\lambda_1^{x,y}(\mathbf q) & 0 & \lambda_1^z(\mathbf q) & 0 \\
	0 & \lambda_1^z(-\mathbf q) & 0 & \lambda_1^{x,y}(-\mathbf q) \\
	\lambda_1^z(\mathbf q) & 0 & \lambda_1^{x,y}(\mathbf q) & 0
	\end{pmatrix},
\end{align}
with
\begin{align}
	\lambda_1^{x,y}(\mathbf q) & = \sum_{\gamma=x,y}  \mathrm e^{\iu\mathbf q\cdot\boldsymbol\delta_\gamma} \Biggl\{
	\frac{K}{2} \left[(\vec e \cdot \vec e_\gamma) - \iu \left((\vec n \times \vec e) \cdot \vec e_\gamma\right) \right]^2
	\nonumber \\ & \quad
	+ \Gamma \Bigl[(\vec e \cdot \vec e_\alpha)(\vec e \cdot \vec e_\beta) - ((\vec n \times \vec e) \cdot \vec e_\alpha) \left((\vec n \times \vec e) \cdot \vec e_\beta\right) 
	\nonumber \\ & \quad
		- \iu (\vec e \cdot \vec e_\alpha)\left((\vec n \times \vec e) \cdot \vec e_\beta\right) - \iu (\vec e \cdot \vec e_\beta)\left((\vec n \times \vec e) \cdot \vec e_\alpha\right) 
	\Bigr]
	\Biggr\}, \displaybreak[2] \\
%\end{align}
%%
%and
%%
%\begin{align}
%
	\lambda_1^{z}(\mathbf q) & = \mathrm e^{\iu\mathbf q\cdot\boldsymbol\delta_z} \Biggl\{
	\frac{K}{2} \left[(\vec e \cdot \vec e_z) - \iu \left((\vec n \times \vec e) \cdot \vec e_z\right) \right]^2
	\nonumber \\ & \quad
	+ \Gamma \Bigl[(\vec e \cdot \vec e_x)(\vec e \cdot \vec e_y) - ((\vec n \times \vec e) \cdot \vec e_x) \left((\vec n \times \vec e) \cdot \vec e_y\right) 
	\nonumber \\ & \quad
		- \iu (\vec e \cdot \vec e_x)\left((\vec n \times \vec e) \cdot \vec e_y\right) - \iu (\vec e \cdot \vec e_y)\left((\vec n \times \vec e) \cdot \vec e_x\right) 
	\Bigr]
	\Biggr\}.
\end{align}
Again, for the relevant high-symmetry wavevectors $\mathbf q$, the eigenvalue equation \eqref{eq:eigenvalue} can be solved in closed form, leading to the analytical formulae displayed in Eqs.~\eqref{eq:omega-min-C2m}, \eqref{eq:omega-max-C2m}, and \eqref{eq:hcC2m}.

%%%%%%%%%%%%%%%%%%%%%%%%%%%%%%%%%%%%%%%%%%%%%%%%%%%%%%%%%%%%%%%%%%%%
%%%%%%%%%%%%%%%%%%%%%%%%%%%%%%%%%%%%%%%%%%%%%%%%%%%%%%%%%%%%%%%%%%%%


\begin{thebibliography}{99}
\bibitem{note3} In the $R\bar3$ structure, the mode dispersion along $\mathbf q = (0,0,L)$ for $\vec h \parallel (1,-2,0)$ agrees with that for $\vec h \parallel (1,0,0)$. Consequently, in Figs.~\ref{fig:disp1_rb3} and \ref{fig:disp2_rb3}, respectively, the data displayed in panels (e) and (h) are identical.
\end{thebibliography}
\end{document}